\documentclass{iopart}
\usepackage{iopams,epsfig}

\usepackage{times}

\def\tfrac#1#2{{\textstyle {#1 \over #2}}}%

\newcommand{\sect}[1]{\setcounter{equation}{0}\section{#1}}
\newcommand{\subsect}[1]{\subsection{#1}}

 \newcommand{\wt}{\widetilde }
  \newcommand{\wh}{\widehat }
\newcommand{\ra}{\rangle }
\newcommand{\la}{\langle }
\newcommand{\cH}{{\cal H} }
\newcommand{\cT}{{\Bbb T} }
\newcommand{\cD}{{\cal D} }
\newcommand{\cQ}{{\cal Q} }
\newcommand{\cE}{{\cal E} }
\newcommand{\cL}{{\cal L} }
\newcommand{\cA}{{\cal A} }
\newcommand{\cB}{{\cal B} }
\newcommand{\cU}{{\cal U} }
\newcommand{\cV}{{\cal V} }
\newcommand{\cG}{{\cal G} }
\newcommand{\cR}{{\cal R} }
\newcommand{\cP}{{\cal P} }
\newcommand{\cI}{{\cal I} }
\newcommand{\cS}{{\cal S} }
\newcommand{\dg}{\dagger }
\begin{document}

\title[Intertwining technique for a system of 
 difference Schr\"odinger equations]{Intertwining technique for a system of 
 difference Schr\"odinger equations and  new 
 exactly solvable multichannel potentials}

\author{
L M Nieto$\dag$, B F
Samsonov$\dag\ddag$ and A A Suzko$\S$
}

\address{
\dag\ Departamento de F\'{\i}sica Te\'orica, Universidad de
Valladolid,  47005 Valladolid, Spain}

\address{
$\ddag$\ Department of Quantum Field Theory, Tomsk State
University, 36 Lenin Ave.,
634050 Tomsk, Russia}

\address{
$\S$\ Joint Institute of Nuclear Technology and Energetics,
 National Academy of Sciences of Belarus, Minsk, Belarus}

\ead{\mailto{luismi@metodos.fam.cie.uva.es},
\mailto{samsonov@phys.tsu.ru} and \mailto{suzko@jinr.ru}
 }

\begin{abstract}
The intertwining
operator technique is applied to difference Schr\"odinger
equations with operator-valued coefficients. It is shown that
these equations appear naturally when a discrete basis is used for
solving a multichannel Schr\"odinger equation. New families of
exactly solvable multichannel Hamiltonians are found. 

\end{abstract}

 \submitto{\JPA}
 {\small{\today}}

\medskip


\medskip
\medskip

\textbf{Corresponding Author}:

L M Nieto

Departamento de F\'\i sica Te\'orica

Universidad de Valladolid

47005 Valladolid, Spain

Tel.: 34 983 42 37 54

Fax: 34 983 42 30 13

E-mail: {\it luismi@metodos.fam.cie.uva.es\/}

\newpage


\sect{Introduction}

The method for finding approximate solutions to the Schr\"odinger
equation based on a matrix representation for the Hamiltonian is
widely used in nonrelativistic quantum mechanics
\cite{QQB}. Its rationale is the intuitive idea that the larger a
finite basis used in particular calculations, the more exact are
the results that may be achieved, and the ``most exact" solution
of the problem corresponds to the entire (infinite) basis set. Of
course, there is now a rigorous justification for this physically
intuitive idea within functional analysis \cite{Neumann},
but we would like to stress here that up to now only a few cases
are known where one is able to manipulate a whole Fourier series
expansion and get the exact solution to the problem. In
particular, the well-known $J$-matrix method \cite{J} (although it
has recently been significantly modified \cite{VBA}), uses a
special truncation procedure for modelling a real scattering
process involving interacting  particles.
Till now
this method uses mainly
 the possibility of getting exact solutions to
the simplest quantum mechanical problem, namely, solving the free
particle Schr\"odinger equation expressed in a special basis as
the eigenvalue problem for a tridiagonal (Jacobi) matrix.

Recently \cite{1} it has been shown that an interaction can be
included in such a model without affecting the tridiagonal form of
the  Hamiltonian, a fact that can significantly enlarge the
range of possible applications for the $J$-matrix method. In this
paper we generalize the results of \cite{1} to the case of a
matrix Schr\"odinger equation, obtaining new
exactly solvable multichannel potentials and, at the same time,
opening the way for much wider applications of the $J$-matrix method.
 In
particular, we will show that there exist exactly solvable potentials
whose matrix representation has the form of a  tridiagonal matrix whose non-zero entries
are not numbers, but matrices.

The paper is organized as follows. In the next Section we show how
an eigenvalue problem for a Hamiltonian possessing both external
and internal degrees of freedom (i.e. a multichannel
Hamiltonian), may be reduced to the eigenvalue problem for a
difference operator having operator-valued
coefficients. We introduce a special space of operator-valued sequences
$\cT \ell^2$, which is the counterpart of the Hilbert space $\ell^2$. In
this space we define an intertwining (or transformation) operator and
its adjoint which relate solutions to two different operator-valued
discrete eigenvalue problems. In Section 3 we give a solution to the
intertwining relation, specifying the transformation operator as well as
the transformed Hamiltonian.  We also show that in the particular case of
a two-dimensional space for an internal degree of freedom, our
formulas
acquire a simplest, though nontrivial, form and
we are able to establish a one-to-one correspondence between the
spaces of solutions of the initial and transformed equations.
 Section 4 is devoted to applying our method to finding a wide class of
new exactly solvable two-channel Hamiltonians whose matrix representation
has  infinite block-tridiagonal form. In the last Section we  outline a
possible continuation of this work.

\sect{Preliminaries}

We will show here how the usual eigenvalue problem for
a Hamiltonian with a composite interaction may be reduced to an
eigenvalue problem for a difference operator having operator-valued (or
matrix-valued) coefficients. 

To begin with, let us consider a time-independent Hamiltonian
$\cH_0$ describing a composite system possessing both external and
internal degrees of freedom. The external degrees of freedom are
related to the motion of the system, considered as a material
point, interacting with time-independent external fields. We shall
deal only with stationary states and describe this motion with the
help of a separable Hilbert space $H$ which, in particular, may be
chosen as the space $L^2(\Bbb R)$. Internal degrees of freedom, related
to the system's structure (spin, colour, charm, \ldots ), will be
associated with an $N$-dimensional linear space ${\Bbb C}^N$. A typical
example is a multichannel quantum system \cite{Newton}. Thus, the
Hamiltonian $\cH_0$ acts in the Hilbert space ${\Bbb H}= H\otimes
{\Bbb C}^N$ and we consider  the eigenvalue problem
\begin{equation}
\cH_0|\Psi \rangle =\cE|\Psi \rangle \,, \quad |\Psi\rangle \in
{\Bbb H} \,. \label{H0}
\end{equation}

Let \{$|n\rangle $, $n=0,1,2,\ldots \}$ be an orthonormal basis in
$H$ and $\{ e_\alpha,\, \alpha=1,\dots,N \}$ be an orthonormal
basis in ${\Bbb C}^N$, so that
 the set
 $\{ |n,\alpha \rangle \equiv |n\rangle  e_\alpha = e_\alpha|n\rangle   \}$
 is a basis in the space $\Bbb H$.
Hence, denoting the Fourier coefficients of an element $|\Psi
\rangle \in \Bbb H$   by $\psi_{n,\,\alpha}$ in the basis
$|n,\alpha \rangle$, one has
\begin{equation}
|\Psi \rangle =\sum_{n,\,\alpha } \psi_{n,\,\alpha}\, |n,\alpha \rangle
=\sum_{n=0}^\infty \Psi_n \, |n\rangle 
\label{develop}
\end{equation}
where $\Psi_n=\sum_{\alpha =1}^N \psi_{n,\,\alpha} e_\alpha$
denotes a column-vector of ${\Bbb C}^N$.
Since the right hand side of the
equation (\ref{develop}) looks like a development of a vector $|\Psi
\rangle$ over the
basis $\{|n\rangle\}$, the quantities $\{ \Psi_n\}$ may be considered as the
``coordinates" of
the ket $|\Psi \ra $.
Following this approach, we can consider the  elements $|\Psi \ra$
of $\Bbb H$,
 as a set of elements endowed with
 a different structure: it is not a
vector space anymore, since its underlying space ${\Bbb C}^N$ is
neither a field nor a ring (only the addition of this type of
objects is defined, but not a multiplication by the elements of
${\Bbb C}^N)$;
 rather, it can be treated as a {\em module over the linear space}
 ${\Bbb C}^N$, which does not correspond to an usual algebraic
construction of
 a module over a ring or a field
\cite{vdw}. We shall denote the set $\{ |\Psi \ra \}$ with this
new structure by $\cT $. Although the kets $\{ |n\ra \}$ are not
elements of $\cT$, we will call them a ``basis" in the sense that
$\forall |\Psi\ra \in \cT$ there exists in ${\Bbb C}^N$ a set
$\{ \Psi_n \}$ such that  (\ref{develop}) is valid, in the sense that
it corresponds to a convergent Fourier series in ${\Bbb H}$. Note
also that no notion of linear dependence or independence can be
introduced in $\cT$, since it is not a linear space, but a module.
Nevertheless, one can introduce an inner product  as a mapping
$\cT \times \cT\to \Bbb C$ inherited from the inner product in
$\Bbb H $, as follows:
\begin{equation}\label{IP}
\langle \Psi|\Phi \rangle :=\sum_{n=0}^\infty \Psi_n^{\,\dagger}\Phi_n =
\sum_{n=0}^\infty \  \sum_{\alpha =1}^N \psi_{n,\,\alpha
}^*\varphi_{n,\,\alpha } ,
\quad \forall \,  |\Psi\ra ,|\Phi\ra\in\cT .
\end{equation}
Here, $\Psi _n^{\,\dagger}$ is the row-vector obtained by
transposition and conjugation of the column-vector $\Psi _n$, and
$\Psi_n^{\,\dagger}\Phi_n$ denotes the inner product of two
elements of  ${\Bbb C}^N$. By simplycity, we  use the same notation
$\langle\cdot |\cdot \rangle $  for the inner product in  $H$ and in
$\cT$.
In terms of this inner product, one can define a distance between
$|\Psi\ra$ and $|\Phi\ra$ as usual:
$
d(\Psi,\Phi)=[\la \Psi -\Phi|\Psi -\Phi\ra ]^{1/2},
$
which converts
 $\cT $ to a metric space. As this metric generates a
(strong) topology in $\cT$, all the topological notions can
be used for studying it, and it can be proved
to be a complete metric space. From this point of
view the right hand side in (\ref{develop}) is a Fourier series
convergent in the strong topology  in $\cT$. The fact that $\cT$
is not a Hilbert space  causes no difficulty from the physical
point of view, since the underlying space $\Bbb H$ is a Hilbert
space, and we introduce $\cT$ only as a convenient tool.

In the sequel we shall impose the following form for the action of the 
operator $\cH_0$ introduced in equation (\ref{H0})  on the
basis
$|n,\alpha\rangle $:
\begin{equation}\label{h0na}
\cH_0|n,\alpha\rangle =\sum_{\beta=1}^N [ D_{n,\alpha ,\beta}|n-1
,\beta\rangle +
D_{n+1 ,\alpha ,\beta}|n+1 , \beta \rangle +Q_{n,\alpha
,\beta}|n,\beta\rangle ] .
\end{equation}
This type of action  corresponds to a tridiagonal matrix whose non-zero
entries are
$N$-dimensional square matrices. The coefficients $D_{n,\alpha
,\beta}$ and $Q_{n,\alpha ,\beta}$  must satisfy $D_{n,\,\alpha
,\,\beta}=D^*_{n,\,\beta ,\,\alpha},
Q_{n,\,\alpha,\,\beta}=Q^*_{n,\,\beta ,\,\alpha}$ to ensure the
Hermitian character of $\cH_0$.

If for a fixed value of $n$ we consider the $N\times N$ matrices
with entries $D_{n,\,\alpha ,\,\beta}$ and $Q_{n,\,\alpha ,\,\beta}$ as
a matrix representation of the self-adjoint operators
$\cD_n=\cD^\dagger_n$ and $\cQ_n=\cQ^\dagger_n$, whose action on
${\Bbb C}^N$ is
\begin{equation}
\cD_ne_\alpha:= \sum_{\beta=1}^N D_{n,\,\alpha ,\,\beta}\, e_\beta \,,
\qquad
\cQ_ne_\alpha:= \sum_{\beta=1}^N Q_{n,\,\alpha ,\,\beta}\, e_\beta \,,
\end{equation}
then, according to (\ref{h0na}),
the action of the Hamiltonian $\cH_0$ on the basis vector
$|n,\alpha\rangle $ takes the form of a three-term relation
with operator-valued coefficients
$$
\cH_0|n,\alpha\rangle =\cD_{n}|n-1 ,\alpha\rangle +
\cD_{n+1 }|n+1 , \alpha
\rangle +\cQ_{n}|n,\alpha  \rangle \, , \quad
\cD_0 |n,\alpha \rangle=0,\, \forall n,\,
\forall \alpha.
$$
We shall allow the operator $\cH_0$ to be unbounded
in ${\Bbb H}$, and to be defined on an appropriate dense set
${\Bbb D}\subset \Bbb H$, which is supposed to be invariant under the
action of
$\cH_0$. Then, $\forall\, |\Psi\rangle \in {\Bbb
D}$ the following Fourier series converges in $\Bbb H$
\begin{equation} \label{FSH}
\cH_0|\Psi \rangle = \sum_{n=0}^\infty \  \sum_{\alpha =1}^N
\ [\,\psi_{n-1,\alpha}
\cD_{n}+\psi_{n+1,\alpha}\cD_{n+1}+\psi_{n,\alpha}\cQ_{n}\,]
|n,\alpha\rangle .
\end{equation}
With the help of this expression and
(\ref{develop}) we can define the action of $\cH_0$ in $\cT$ to be:
\begin{equation}\label{h0psin}
\begin{array}{l}
\displaystyle\cH_0|\Psi \rangle \! =\! \sum_{n=0}^\infty \ (\cH_0\Psi )_n\
|n\ra, \\ [3ex]
(\cH_0\Psi )_{n} :=
\cD_{n+1}\Psi_{n+1}+ \cD_{n}\Psi_{n-1}+\cQ_n\Psi_n .
\end{array}
\end{equation}
Note that the right hand side in the first equation is a convergent
Fourier series in $\Bbb T$ since it just corresponds to the
convergent Fourier series (\ref{FSH}) in $\Bbb H$. Moreover, $\cH_0$, as
an operator acting in $\Bbb T$, has a well-defined domain of definition
$\widetilde {\Bbb D}\subset \Bbb T$ consisting of all elements of $\Bbb
D$  considered as elements of $\Bbb T$. We note that since $\Bbb D$ is
invariant under the action of $\cH_0$, then $\widetilde {\Bbb D}$
is also invariant under the action of $\cH_0$ acting on $\Bbb T$.

The eigenvectors of $\cH_0$ are obtained from the
finite-difference eigenvalue problem
\begin{equation}\label{IMEP}
\cD_{n+1}\Psi_{n+1}+ \cD_{n}\Psi_{n-1}+\cQ_n\Psi_n = \cE\Psi_n 
\end{equation}
with operator-valued coefficients
(cf. \cite{Berezanskii}). We do not indicate explicitly the dependence of
$\Psi_n$  on $\cE$ in
order  to avoid a cumbersome notation.

Let us suppose now that some solutions of the eigenvalue problem
for $\cH_0$ are known and we want to get a solution of the
eigenvalue problem for another Hamiltonian $\cH_1$.
According to the usual strategy of
intertwining operators \cite{BS}, one can try to solve this
problem by finding an intertwiner (or transformation operator)
$\cL$ defined by
\begin{equation}
\label{intertw} \cL\cH_0=\cH_1\cL 
\end{equation}
where we consider all operators acting on the same domain 
$\widetilde{\Bbb D}\subset \Bbb T$. Once $\cL$ is found, the eigenkets
$|\widetilde\Psi\rangle $ of $\cH_1$ are obtained
by applying $\cL$ to the eigenkets $|\Psi \rangle $ of $\cH_0$:
$|\widetilde\Psi\rangle =\cL|\Psi \rangle $. Here, both $|\Psi \rangle $
and $|\wt\Psi \rangle $ correspond to the same eigenvalue $\cE$.

We  consider now a particular ansatz for $\cL$, by assuming that
it acts on the basis element  $|n,\alpha\rangle $ of the space $\Bbb
H$  as follows:
\begin{equation}
\label{Ln}
\cL |n,\alpha\rangle =\sum_{\beta=1}^N \, [\,  A_{n,\,\alpha ,\,\beta}\,
|n-1,\beta\rangle
+B_{n,\,\alpha ,\,\beta}\, |n,\beta\rangle \,] 
\end{equation}
with $A_{0,\,\alpha ,\,\beta}\equiv 0$. We easily
find that the action of $\cL$ on any $|\Psi \ra$ from ${\Bbb D}
\subset \Bbb H$ is
\begin{equation}\label{Lpsin}
\begin{array}{l}
\displaystyle
\cL |\Psi\rangle  = |\widetilde\Psi\rangle
=\sum_{n=0}^\infty
\widetilde\Psi_n\, |n\rangle
\,,\\ [3ex]
\widetilde\Psi_n  \equiv
(\cL\Psi)_n  := \cA_{n+1}\Psi_{n+1}+\cB_n\Psi_n.
\end{array}
\end{equation}
Here $\cA_n$ and $\cB_n$ are operators acting on ${\Bbb C}^N$, which in
the basis $\{e_\alpha\}$ are represented by   matrices with entries
$A_{n,\,\alpha ,\,\beta}$ and $B_{n,\,\alpha ,\,\beta}$. For completeness,
we remark that $\cA_0=0$.

It should be mentioned that the second relations of (\ref{h0psin}) and
(\ref{Lpsin}) define the action of $\cH_0$ and $\cL$ on the space
of vector-sequences $\left\{\Psi_n\right\}$, which we shall denote  by
$\cT\ell ^{\,2}$, since it is just a counterpart of the
Hilbert space $\ell ^{\,2}$. To be precise, we should remark that
these are not really the same operators as the ones defined above, since
they  act on different spaces. Nevertheless, we will use the same
notations for them and for the inner product in $\cT\ell^{\,2}$,
believing that this will not cause any trouble to the reader. The
inner product in $\cT\ell^{\,2}$ is defined as usual by
(\ref{IP}),
where now $\Psi:=\left\{\Psi_{n}\right\},\,
\Phi:=\left\{\Phi_{n}\right\}$ are elements of ${\Bbb T}\ell^2$.
Once we have an inner product in $\cT\ell^{\,2}$, we
are able to determine the adjoint  $\cL^\dagger$ of an  operator $\cL$,
defined as usually by the relation
$\la\Psi|\cL\Phi\ra = \la\cL^\dagger\Psi|\Phi\ra$,
$\Psi ,\, \Phi\in\cT\ell^2$.
We shall consider $\cL^\dagger$ as acting on finite elements which form a
dense set in
$\cT\ell^{\,2}$.
 Using
equations (\ref{Lpsin}), we evaluate $\la\Psi|\cL \Phi\ra$
and find the action of $\cL^\dagger$
\begin{equation}
\label{Ldag}
(\cL^\dagger\Psi)_n=\cA^{\,\dagger}_n\Psi_{n-1}+ \cB^{\,\dagger}_n\Psi_n
\end{equation}
where $\cA^{\,\dagger}_n$ and $\cB^{\,\dagger}_n$ are the
operators adjoint to $\cA_n$ and $\cB_n$, respectively.

\sect{Discrete-matrix intertwining technique}

\subsect{Intertwining technique in the space $\cT\ell ^{\,2}$}

The second
relations of (\ref{h0psin}) and (\ref{Lpsin}) define the action of
$\cH_0$ and $\cL$ on any finite element $\{\Psi_n\}$ of  $\cT\ell
^{\,2}$,  meaning that the sequence $\{\Psi_n\}$ has a finite
length. For the product $\cL\cH_0$ one gets
\begin{eqnarray}
\cL(\cH_0 \Psi )_n &=&
\cA_{n+1}(\cD_{n+2}\Psi_{n+2}+\cQ_{n+1}\Psi_{n+1}+\cD_{n+1}\Psi_{n})
\nonumber  \\
 && +\cB_n(\cD_{n+1}\Psi_{n+1}+\cQ_{n}\Psi_{n}+\cD_{n}\Psi_{n-1})\,.
\end{eqnarray}
A similar expression is obtained
for  $\cH_1\cL$.
Then the
intertwining relation (\ref{intertw}) implies
\begin{eqnarray}
& &\cA_{n}\cD_{n+1}=\widetilde{\cD}_{n}\cA_{n+1}  \label{sys1}\\ [1ex]
& &\cB_n\cD_n=\widetilde{\cD}_n\cB_{n-1}\label{sys2} \\ [1ex]
& &\cA_{n+1}\cQ_{n+1}+\cB_n \cD_{n+1}=
\widetilde{\cD}_{n+1}\cB_{n+1}+\widetilde{\cQ}_n\cA_{n+1} \label{sys3}\\
[1ex]
& &\cA_{n+1}\cD_{n+1}+\cB_n\cQ_n=
\widetilde{\cD}_n \cA_n+\widetilde{\cQ}_n\cB_n\,.
\label{sys4}
\end{eqnarray}
When solving these equations (the unknowns being the operators
$\widetilde{\cD}_n, \widetilde{\cQ}_n, \cA_n, \cB_n$, and the data
${\cD}_n, {\cQ}_n$) we should take into account that $\cA_0=\cD_0=0$. By
assuming the operators $\cA_n, \cB_n$ to be invertible, we easily find
from (\ref{sys2})
\begin{equation}
\label{wt}
\wt \cD_n=\cB_n\cD_n(\cB_{n-1})^{-1}
\end{equation}
which being placed to (\ref{sys1}) gives us
\begin{equation}\label{2a}
\cA_n\cD_{n+1} (\cA_{n+1})^{-1}=\cB_n\cD_n (\cB_{n-1})^{-1}.
\end{equation}
Similarly, from (\ref{sys4}) we have
\begin{equation}\label{wtqn}
\wt \cQ_n=\cA_{n+1}\cD_{n+1}(\cB_n)^{-1}+\cB_n\cQ_n(\cB_n)^{-1}-\wt\cD_n
\cA_n(\cB_n)^{-1}
\end{equation}
which together with (\ref{sys3}) yields
\begin{eqnarray}\label{3a}
\cA_{n+1}\cQ_{n+1}+\cB_n\cD_{n+1}
&=&
\wt \cD_{n+1}\cB_{n+1}+\cA_{n+1}\cD_{n+1} (\cB_{n})^{-1}\cA_{n+1}
 \\
 &&+   \cB_n\cQ_n (\cB_n)^{-1}\cA_{n+1}-\wt \cD_n \cA_n (\cB_n)^{-1}
\cA_{n+1}\,. 
\nonumber
\end{eqnarray}
To solve this equation we proceed to eliminate the variable
$\cB_n$ by introducing a new auxiliary variable (actually,
operator) $\sigma_n$ defined by
\begin{equation}
\label{Bsigma}
\cB_n=\cA_{n+1}\sigma _n\,.
\end{equation}
In equation (\ref{3a}) we replace
 $\wt \cD_n$ by its expression (\ref{wt}),
 multiply the result from the left by $(\cA_{n+1})^{-1}$
and use (\ref{2a}) and (\ref{Bsigma})  to rearrange the
first term on the right hand side to get a nonlinear finite-difference
equation where the only unknown is $\sigma_n$:
\begin{equation}
\label{differenceeq}
\cQ_{n+1}-\cD_{n+2}\sigma_{n+1}-\cD_{n+1}\sigma_n^{-1}=
\sigma_n(\cQ_n-\cD_{n+1}\sigma_n-\cD_n\sigma_{n-1}^{-1})\sigma_n^{-1}\,.
\end{equation}
This  equation having operator-valued
coefficients may be interpreted as a matrix finite-difference
counterpart of the derivative of a Riccati  equation. It can be
linearized and ``integrated" by introducing the new variable $\cU_n$
defined by
\begin{equation}
\label{U}
\sigma_n=-\cU_{n+1}\cU_n^{-1}\,.
\end{equation}
Using it in (\ref{differenceeq}) and
simplifying, we get the difference equation $G_{n+1}=G_n$, with
\begin{equation}\label{zzpaf}
G_n =\cU_{n}^{-1} [ \cQ_{n}+\cD_{n+1}\cU_{n+1} \cU_{n}^{-1} +
\cD_{n} \cU_{n-1}\cU_{n}^{-1} ] \cU_{n} \,,
\end{equation}
whose solution is $G_n=\Lambda$. The matrix
$\Lambda$  plays the role of  an integration constant. This leads us to
an equation for $\cU_n$
\begin{equation}\label{Ueq}
\cD_{n+1}\cU_{n+1}+\cD_{n}\cU_{n-1}+\cQ_n\cU_n=\cU_n\Lambda
\end{equation}
which is identical to the eigenvalue problem (\ref{IMEP}), except for
the fact that the vector-valued variable $\Psi_n$ is replaced by
the matrix-valued variable $\cU_n$ and the scalar parameter $\cE$
is replaced by the matrix $\Lambda $. We shall now show  how
solutions of (\ref{Ueq}) may be obtained from solutions of
(\ref{IMEP}). Let us choose $\Lambda$ to be diagonal: $\Lambda
=\mbox{diag}(\lambda_1,\ldots \lambda_N)$. In fact, this is not a
restriction, since this matrix is Hermitian (this follows from the
Hermicity of the matrices $\cD $ and $\cQ $) and therefore it can
be diagonalized by an appropriate unitary transformation. Denote
the columns of the matrix $\cU_n$ by $U_{1,n},\ldots ,U_{N,n}$. If these
vectors are solutions to the initial  eigenvalue problem
\begin{equation}
\label{initialeigenvalueproblem}
\cD_{n+1}U_{i,n+1}+\cD_nU_{i,n-1}+\cQ_nU_{i,n}=\lambda_iU_{i,n}\,,\quad
i=1,\ldots ,N
\end{equation}
then the matrix $\cU_n$ is a solution to equation (\ref{Ueq}).
Once the matrices $\cU_n$ are given, one can evaluate $\cB_n$
using (\ref{U}) and (\ref{Bsigma}), provided the matrices $\cA_n$
are known.

Now, we shall obtain an expression for $\cA_n$. In order to do that, we have
to solve
equation (\ref{2a}), where $\cB_n$ are given in (\ref{Bsigma}), that is
\begin{equation}
\label{EqA}
\cA_n\cD_{n+1}\cA_{n+1}^{-1}= \cA_{n+1}R_{n}\cA_{n}^{-1}\,.
\end{equation}
Note that the quantities
\begin{equation}
\label{Rn}
R_n=\sigma_n\cD_n\sigma_{n-1}^{-1}, \quad n=1,2,\dots
\end{equation}
are known, because they depend only on the initial data $\cD_{n}$
and the matrices $\sigma _n$ (or $\cU_n$), which are given. Using very
simple algebra,  one obtains from (\ref{2a})
\begin{equation}\label{AR}
\cD_{n+1}R_n=(\cA_n^{-1}\cA_{n+1}R_n)^2\,,\quad
R_n\cD_{n+1}=(\cA_{n+1}^{-1}\cA_{n}\cD_{n+1})^2\,.
\end{equation}
It is a simple exercise to see that these equations are
compatible, which means that equation (\ref{EqA}) has a nontrivial
solution. Hence, using for instance the first one, we get
\begin{equation}
\cA_{n+1}=\cA_n(\cD_{n+1}R_n)^{1/2}R_n^{-1}\,.
\end{equation}
From this recursion relation  we find all the matrices $\cA_n,
n=2,3,\dots,$ in terms of $\cA_1$:
$$
\cA_n=\cA_1\left[ (\cD_2R_1)^{1/2}R_1^{-1}\right] \left[
(\cD_3R_2)^{1/2}R_2^{-1}\right] \cdots \left[
(\cD_nR_{n-1})^{1/2}R_{n-1}^{-1}\right]    .
$$
Once
$\cA_n$ and $\cB_n$ are found, from 
(\ref{sys1})--(\ref{sys2}) we have two equivalent expressions for
$\wt\cD_n$:
\begin{equation}
\label{tDn}
\wt\cD_n=\cA_{n+1}\sigma_n\cD_n\sigma_{n-1}^{-1}\cA_n^{-1}=
\cA_n\cD_{n+1}\cA_{n+1}^{-1}\,.
\end{equation}
To find $\wt\cQ_n$ we use  (\ref{sys4}), which yields
\begin{equation}\label{tQn}
\wt\cQ_n=\cA_{n+1}(
\cD_{n+1}+\sigma_n\cQ_n-\sigma_n\cD_n\sigma_{n-1}^{-1} )\sigma
_n^{-1}\cA_{n+1}^{-1}\,.
\end{equation}

It is necessary to mention that in finding the matrices $\wt\cD_n$
and $\wt\cQ_n$ we never required them to be self-adjoint.
Therefore, our solution can provide us with both self-adjoint and
non-self-adjoint matrices. The conditions $\wt\cD_n
=\wt\cD_n^{\,\dagger }$ and $\wt\cQ_n =\wt\cQ_n^{\,\dagger }$ are
really restrictions on the transformation function $\cU_n$.

Note that the quantities $\{\cU_n\}$ uniquely define both the
transformation operator $\cal L$ and the new ``potentials" $\wt\cD_n$ and
$\wt\cQ_n$. Therefore, we shall call  $\cU=\{\cU_n\}$ the {\it
transformation function}.

\subsect{A particular case}

In this Section we shall consider a particular nontrivial case
corresponding to $N=2$,
where general formulas established in the previous Section take
the simplest form. Moreover,
we shall show that
for this case
one can establish one-to-one-correspondence between the spaces of
solutions of the initial and transformed equations and investigate
the factorization properties of transformation operators.

Let us put $N=2$ and choose
the matrices $\cD_n$ and $\cQ_n$ to be multiples of the $2\times
2$ identity matrix $\cI$: $\cD_n=d_n \cI$, $\cQ_n=q_n \cI$. Let
$u_{jn}$ be real
 solutions to the scalar difference equation
\begin{equation}\label{one_chan}
d_{n+1}u_{j,n+1}+d_nu_{j,n-1}+q_nu_{j,n}=\lambda_ju_{j,n} ,\quad j=1,2\,.
\end{equation}
Then, the matrix-function
\begin{equation}
\label{U2n}
\cU_n=\left(
\begin{array}{cr}
u_{1,n}&-u_{2,n}\\
u_{1,n}&u_{2,n}
\end{array}
\right)
\end{equation}
is a solution to the initial eigenvalue problem
(\ref{Ueq}), with $\Lambda
=\mbox{diag}(\lambda_1,\lambda_2)$.
We will show  that it is
suitable to be used as the transformation function in our algorithm,
and that it will lead us to a nontrivial exactly solvable
matrix-difference eigenvalue problem. After finding the inverse
function
\begin{equation}\label{Un1}
\cU_n^{-1}=\Delta_n^{-1}
\left(
\begin{array}{rc}
u_{2,n}&u_{2,n}\\
-u_{1,n}&u_{1,n}
\end{array}
\right),
\end{equation}
where $\Delta_n=2\, u_{1,n}\, u_{2,n}$, we get from (\ref{U})
\begin{equation}
\label{sig}
\sigma_n=-\cU_{n+1}\cU_n^{-1}=-\Delta_n^{-1}
\left(
\begin{array}{cr}
\omega_{n}&\pi_{n}\\
\pi_{n}&\omega_{n}
\end{array}
\right)
\end{equation}
with
\begin{equation}
\label{omegapi}
\omega_n=u_{1,n+1}u_{2,n}+u_{1,n}u_{2,n+1} \  {\rm and}\ 
\pi_n=u_{1,n+1}u_{2,n}-u_{1,n}u_{2,n+1}\,.
\end{equation}
Two remarkable properties of the matrices (\ref{sig}) should be
mentioned:
\begin{itemize}
\item[(i)] From equations (\ref{sig}) and  (\ref{omegapi}), and
from the real character of $u_{j,k}$,  it is obvious that all the
matrices $\sigma_n$ are Hermitian: $\sigma_n^\dagger=\sigma_n$;
\item[(ii)] They form a commutative group with respect to  matrix
multiplication (this assertion can be trivially proved using
(\ref{sig})).
\end{itemize}
The last fact simplifies  considerably our formulas. For
instance, all the matrices in the expression for $\cA_n$ commute, and
we easily find
\begin{equation}
\label{Anp}
\cA_n= \cD_n^{1/2}\ (\cU_{n-1}\ \cU_n^{-1})^{1/2}
\end{equation}
where $\cA_1$ has been chosen appropriately.
The expression for $\cB_n$ follows from (\ref{Bsigma}):
\begin{equation}
\label{Bnp}
\cB_n=  -\cD_{n+1}^{1/2}\ (\cU_{n+1}\ \cU_{n}^{-1})^{1/2}.
\end{equation}
Using (\ref{tDn}) and (\ref{tQn}) we deduce formulas for
$\wt\cD_n$ and $\wt\cQ_n$:
\begin{equation}\label{wtD}
\left\{
\begin{array}{l}
\wt\cD_n =
(\cD_n\ \cD_{n+1}\ \cU_{n-1}\ \cU_n^{-1}\ \cU_{n+1}\ \cU_n^{-1})^{1/2}
\\ [2ex]
\wt\cQ_n  =
\cQ_n-\cD_{n+1}\ \cU_n\  \cU_{n+1}^{-1}+\cD_n\ \cU_{n-1}\ \cU_n^{-1} \,.
\end{array}
\right.
\end{equation}
 For our particular choice ($\cD_n=d_n \cI$, $\cQ_n=q_n \cI$), we 
get explicit expressions for these matrices in terms of the known
solutions of  (\ref{one_chan}). Indeed, using 
(\ref{U2n})--(\ref{Un1}), and choosing the positive 
square root in the first of equations (\ref{wtD}) we get  
$\wt \cD _n=
 \left(
 \begin{array}{cc}
 a_+ & a_- \\
 a_- & a_+
 \end{array}
 \right)$, where
 \begin{equation}\label{Dmatr}
 a_{\pm}=\frac{\sqrt{d_n\,d_{n+1}}}{2}
 \left[
 \sqrt{\frac{u_{1,n+1}u_{1,n-1}^{\vphantom i}}{u_{1,n}^2}}\,\pm \,
  \sqrt{\frac{u_{2,n+1}u_{2,n-1}^{\vphantom i}}{u_{2,n}^2}}~~
 \right] .
 \end{equation}
Similarly, from the second of equations (\ref{wtD}) one finds $\wt \cQ _n= \cQ_n +
 \left(
 \begin{array}{cc}
 b_+ & b_- \\
 b_- & b_+
 \end{array}
 \right)$, where
 \begin{equation}\label{Qmatr}
 b_{\pm}=\frac{d_n}{2}
 \left[
 \frac{u_{1,n-1}}{u_{1,n}}\,\pm \,
 \frac{u_{2,n-1}}{u_{2,n}}
 \right] -
 \frac{d_{n+1}}{2}
 \left[
 \frac{u_{1,n}}{u_{1,n+1}}\,\pm \,
 \frac{u_{2,n}}{u_{2,n+1}}
 \right].
 \end{equation}
We notice that for $u_{2,n}=u_{1,n}$ the matrices $\wt\cD_n$
 and $\wt\cQ_n$ become diagonal, with the nonzero elements
 coinciding  with the previously obtained for the
 scalar case \cite{1}.

From  (\ref{Dmatr}) and (\ref{Qmatr}), it is clear that the transformed
quantities are Hermitian provided $a_{\pm}$ and $b_{\pm}$ are real.
 On the other hand, it is easy to show that,
under some reasonable assumptions on $q_n$ and $d_n$, the
operators $\cA_n$ and $\cB_n$ are anti-Hermitian.
Indeed, from (\ref{Ueq}) we have
\begin{equation}\label{Ueq22}
\cD_{n+1}\cU_{n+1} \cU_n^{-1}+\cD_{n}\cU_{n-1}
\cU_n^{-1} =\cU_n\Lambda  \cU_n^{-1} - \cQ_n \, .
\end{equation}
By choosing  the matrix $\Lambda$ appropriately, and provided the
sequence $\{q_n\}$ is bounded from below,  the matrices in both
sides of (\ref{Ueq22}) may be negative definite for all values
of $n$ to get the matrix
$
\cD_{n+1}\sigma_{n} +\cD_{n} \sigma_{n-1}
$
positive definite. This is achieved by taking the operators
$\cD_{n}$, as well as $\sigma_{n}$, to be positive definite
$\forall n$, and the anti-Hermiticity of  $\cA_n$ and $\cB_n$
 follows from
(\ref{Anp}) and (\ref{Bnp}).

Let us now calculate the superposition $\cL^\dg\cL
$ as an operator acting on $\cT\ell ^{\,2}$. Using equations
(\ref{Lpsin}), (\ref{Ldag}), (\ref{Anp}), (\ref{Bnp}), and
the anti-Hermiticity of $\cA_n$ and $\cB_n$ we find
\begin{eqnarray}
 (\cL^\dg\cL\Psi )_n &=&
(\cD_{n+1}\Psi_{n+1}+ \cD_n\Psi_{n-1}+ \cQ_n\Psi_n) \nonumber \\ 
&&-
(\cD_{n+1}\cU_{n+1}+ \cD_n\cU_{n-1}+ \cQ_n\cU_n)\cU_n^{-1}\Psi_n \,.
\end{eqnarray}
The first term on the right hand side is simply the function
$(\cH_0\Psi )_n$ (see equation (\ref{h0psin})). In the
parentheses of the second term we see the action of the same
operator on the transformation function $\cU_n$. Hence, using
(\ref{Ueq}) we transform the right hand side of the previous
equation to the form
$
(\cL^\dg\cL\Psi )_n=
(\cH_0-\cU_n\Lambda \cU_n^{-1})\Psi_n
$.
For the particular case of the transformation function given
in (\ref{U2n}) we easily obtain
\begin{equation}
\cU_n\Lambda \cU_n^{-1}=
\frac 12
\left(
\begin{array}{lr}
\lambda_{1}+\lambda_{2} & \lambda_{1}-\lambda_{2}   \\
\lambda_{1}-\lambda_{2} & \lambda_{1}+\lambda_{2}
\end{array}
\right) =:\wt\Lambda 
\end{equation}
and therefore
\begin{equation}
\label{LdLxx}
\cL^\dg\cL=\cH_0-\wt\Lambda \,.
\end{equation}
We observe here a difference between our Darboux technique
and supersymmetric quantum mechanics usually based on the factorization
$
\cL^{\,\dg}\cL=\cH_0-\Lambda
$.
The factorization constant (in our case $\wt\Lambda$) does not
coincide with an eigenvalue of the transformation function. We
believe that this is a manifestation of the fact that
supersymmetric quantum mechanics based only on the
factorization of the Hamiltonian is a particular case
of the Darboux transformation method, which
in general allows to factorize other symmetry operators.
A similar factorization takes place for
the inverse superposition
$
\cL\cL^{\,\dg}=\cH_1-\wt\Lambda
$
which may be established by applying the operator $\cL$ to formula
(\ref{LdLxx}).

Our results indicate the existence of a one-to-one correspondence
between the solution spaces of the initial and transformed
equations.  To develop  this, we  find  the second solution
$\wh\cU_n$ of  (\ref{Ueq}) with a given value of
$\Lambda$,
$\cD_{n+1}\wh\cU_{n+1}+\cD_{n}\wh\cU_{n-1}+\cQ_n\wh\cU_n=\wh\cU_n\Lambda$.
By eliminating $\cQ_n$ from this equation and using the
adjoint form of  (\ref{Ueq}), we get
$
\cU_n^{\,\dg}\cD_{n+1}\wh\cU_{n+1}- \cU_{n+1}^{\,\dg}\cD_{n+1}\wh\cU_{n}+
\cU_n^{\,\dg}\cD_{n}\wh\cU_{n-1} - \cU_{n-1}^{\,\dg}\cD_{n}\wh\cU_{n} =
\cU_n^{\,\dg}\wh\cU_n\Lambda -\Lambda \cU_n^{\,\dg}\wh\cU_n \,.
$
Because of our particular choice of the transformation function,
the right hand side of the last equation vanishes and the
resulting difference equation can be ``integrated" to yield
$$
\cU_{n-1}^{\,\dg}\cD_{n}\wh\cU_{n} -  \cU_n^{\,\dg}\cD_{n}\wh\cU_{n-1}=
\cU_n^{\,\dg}\cD_{n+1}\wh\cU_{n+1}- \cU_{n+1}^{\,\dg}\cD_{n+1}\wh\cU_{n}=
W_0\,.
$$
From here we see that the matrix $W_0$ plays the role of the Wronskian for
our discrete eigenvalue problem, and we find the recursion relation
for $\wh\cU_{n}$, which being iterated gives
\begin{equation}\label{sec_sol}
 \wh\cU_{n} = (\cU_{0}^\dg)^{-1} \ \cU_n^{\,\dg}\ \wh\cU_0+
 \sum_{k=1}^n\cD_k^{-1} \ (\cU_{k}^\dg)^{-1} \  \cU_n^{\,\dg}\
 (\cU_{k-1}^\dg)^{-1} \ W_0 \,.
\end{equation}
Now, we can act with $\cL$ on this function in order to get a
solution of the transformed equation with eigenvalue $\Lambda$.
After some algebraic manipulation we arrive at
\begin{equation}\label{cSn}
\cS_n\equiv (\cL\ \wh\cU)_n =
\cD_{n+1}^{-1/2}\ (\cU_{n}\, \cU_{n+1}^{-1})^{1/2}\ (\cU_n^{\,\dg})^{-1}
\end{equation}
where the constant matrix $W_0$ has been chosen to be the identity.
We note also that the operator-valued function $\cS_n$ belongs to the
kernel of the operator $\cL^{\,\dg}$, i.e. $\cL^{\,\dg}\cS_n =0$. This
can  be established easily by using formulas (\ref{Ldag}), (\ref{Anp}) and
(\ref{Bnp}). Once $\cS_n$ is found, one can get the second
solution $\wh\cS_n$
 of the transformed equation by using
(\ref{sec_sol}) with appropriate changes. The solution $\wh\cU_n$
also corresponds to the matrix eigenvalue $\Lambda$.

The operator $\cL^{\,\dg}$ satisfies the intertwining relation
 adjoint to (\ref{intertw}). This means that it is a
 transformation operator from solutions of the transformed
 equation to solutions of the initial one.
 Then,  if $E\ne \lambda_{1,2}$, the function
 $\wt\Psi_n=(\cL\Psi )_n$ is a nontrivial solution of the
 transformed equation, whereas
 $\Phi_n=(\cL^{\,\dg}\wt\Psi )_n$ is also a nontrivial solution,
 but to the initial equation. Solutions of the transformed
 equation with eigenvalues $\lambda_{1,2}$ can be found using
  (\ref{sec_sol})--(\ref{cSn}). Hence, we have constructed a
one-to-one
 correspondence  between the solution spaces
 of the initial and transformed eigenvalue problems.

\sect{Application}

We shall use the above developed discrete-matrix Darboux
transformation to generate new exactly solvable two-channel
potentials. Indeed, we will
 show that even
 starting with the simplest case of the two-channel uncoupled
 free particle Hamiltonian, we can  obtain a family of nontrivial coupled
 exactly solvable potentials. In contradistinction to all known
exactly
 solvable potentials, they are represented by infinite block-tridiagonal
 matrices.  The eigenvalue problem for such
matrices is not solvable by usual methods.

Consider the free particle Hamiltonian $h_0=p_x^2$.
Since the momentum operator  $p_x$ is expressible in
terms of the harmonic oscillator ladder operators $a^+=id/dx+ix/2$ and
$a=id/dx-ix/2$, $h_0=(a+a^+)^2/4$ is a quadratic form in $a$ and $a^+$.
Therefore, the action of $h_0$ on the oscillator basis $|n\rangle $,
which in coordinate representation is
\begin{equation}\label{psinH}
\psi_n(x)= \langle x|n\rangle = (-i)^n(n!2^n\sqrt{2\pi })^{-1/2}
e^{-x^2/4}H_n(x/\sqrt 2)
\end{equation}
where $H_n(z)$ are Hermite polynomials, takes the form of a three
term relation
\begin{equation} \label{h0one}
h_0|n\rangle =
d_n|n-2\rangle +
d_{n+2}|n+2\rangle +
q_n|n\rangle,\ 
d_n=\tfrac{\sqrt{n(n-1)}}4,\ 
q_n=\tfrac n2+\tfrac 14.
\end{equation}
To derive (\ref{h0one}) we have used
 $a|n\rangle =\sqrt n|n-1\rangle $ and $a^+|n\rangle
=\sqrt{n+1}|n+1\rangle $.
Let $|\psi_E\rangle $ be
a continuous spectrum eigenket of $h_0$:  $h_0|\psi_E\rangle
=E|\psi_E\rangle $. Then, using the Hermiticity of $h_0$ we get the scalar
 discrete eigenvalue problem
\begin{equation}  \label{in1}
d_n\psi_{n-2}+d_{n+2}\psi_{n+2}+q_n\psi_n=E\psi_n\,.
\end{equation}
 A ``physical"
 solution  $\psi_n=\psi_n(E)=\langle \psi_E|n\rangle, E>0$,
to this equation can be easily obtained since it coincides with the
Fourier image  of the function (\ref{psinH}):
\begin{equation} \label{psinHE}
\psi_n(E)=2(n!2^n\sqrt{2\pi })^{-1/2}e^{-E}H_n(\sqrt {2E}\,)~.
\end{equation}
Consider now an uncoupled two-channel problem with the diagonal matrix
Hamiltonian $\cH_0=h_0\cI$. This is only the  kinetic energy
operator  which
 acts in the Hilbert space ${\Bbb H}=H\otimes {\Bbb C}^{\,2}$,
 with basis $|n,\alpha\rangle =e_\alpha |n\rangle$,
 $e_1=(1,0)^t$, $e_2=(0,1)^t$.
 The action of
 $\cH_0$ on a two-component vector
 $|\Psi\rangle =\sum_{n=0}^\infty \Psi_n|n\rangle $,
  is given by (\ref{h0psin}), where
\begin{equation}
(\cH_0\Psi )_{n}  =
\cD_{n+2}\Psi_{n+2}+ \cD_{n}\Psi_{n-2}+\cQ_n\Psi_n \in {\Bbb C}^{\,2}
\end{equation}
$\cD_n=d_n\cI$ and $\cQ_n=q_n\cI$.
Since the solutions of the one-channel problem
for $h_0$ (\ref{in1}) are known, we know solutions of the two-channel problem
for $\cH_0$. In particular, two-column vectors
$\left\{\Psi_n^+(E)\right\}= \left\{(\psi_n(E),0)^t\right\}$
and
$\left\{\Psi_n^-(E)\right\}= \left\{(0,\psi_n(E))^t\right\}$
with $\psi_n(E)$ given in (\ref{psinHE})
are
solutions to equation
\begin{equation} \label{init2}
\cD_{n+2}\Psi_{n+2}+ \cD_{n}\Psi_{n-2}+\cQ_n\Psi_n =\cE \Psi_n
\end{equation}
 with $\cE=E$. The matrix $\cU_n$ given in (\ref{U2n}),
 with matrix elements
\begin{equation}\label{u1n2}
u_{1n}=(n!\,2^n)^{-1/2}H_n(\sqrt{2\lambda_1})~,\quad
u_{2n}=(n!\,2^n)^{-1/2}H_n(\sqrt{2\lambda_2})
\end{equation}
$\lambda_{1,2}<0 $, is a solution to the matrix eigenvalue problem
\begin{equation}\label{Ueq2}
\cD_{n+2}\cU_{n+2}+\cD_{n}\cU_{n-2}+\cQ_n\cU_n=\cU_n\Lambda 
\end{equation}
with $\Lambda =\mbox{diag}(\lambda_1,\lambda_2)$.
Remark that this two-channel problem is not exactly of
the type considered in Section 2: here the eigenvalue problem involves
shifting by two units the discrete variable $n$.
 Nevertheless, it is not difficult
to see that we can consider here two independent problems,
one for the even and the other for the odd values values of $n$. For the
sake of simplicity we omit these details and give below only the
final results.

Now let the Hamiltonian $\cH_1=\cH_0+\cV$ have the
interaction potential
\begin{equation}\label{Vn}
\cV |n,\alpha\rangle =\cG_{n}|n-2 ,\alpha\rangle + \cG_{n+2 }|n+2 ,
\alpha \rangle + \cR_{n}|n,\alpha  \rangle \,,\quad \alpha =1,2
\end{equation}
with coefficients $\cG_{n}$ and $\cR_{n}$ that are to be
determined. Let $|\widetilde\Psi\rangle$ be an eigenvector of
$\cH_1$, $\cH_1|\widetilde\Psi\rangle=\cE |\widetilde\psi\rangle$,
and $\left\{\right.\wt\Psi_n\left.\right\}$ be its Fourier
coefficients over the basis $|n\ra$. Let these quantities satisfy
the similar discrete matrix eigenvalue problem
\begin{equation} \label{tdnrn}
\wt \cD_{n}\wt\Psi_{n-2}+\wt \cD_{n+2}\wt \Psi_{n+2}+
\wt \cQ_n\wt \Psi_n =
\cE\wt\Psi_n  \,.
\end{equation}
Suppose also that the unspecified quantities $\cG_{n}$ and
$\cR_{n}$ are related to $\wt \cD_{n}$ and $\wt \cQ_n$ by
\begin{equation}\label{dnfi}
\widetilde \cD_n=\cG_n+
 \tfrac 14\sqrt{n(n-1)}\,\cI\,,
 \quad
 \widetilde \cQ_n=\cR_n+(\tfrac n2+\tfrac 14)\,\cI\,.
\end{equation}
Consider the subclass of potentials (\ref{Vn}) for which equation
(\ref{tdnrn}) coincides with the Darboux transform of
the initial eigenvalue problem (\ref{init2})
 and express the transformation function $\cU_n$ in
the form (\ref{U2n}) with the entries given in (\ref{u1n2}).
All these assumptions are satisfied if the
functions $\widetilde \cD_n$ and $\widetilde \cQ_n$
have the form
\begin{equation}\label{wtDp}
\left\{
\begin{array}{l}
\wt\cD_n=
[\cD_n\cD_{n+2}\cU_{n-2}\cU_n^{-1}\cU_{n+2}\cU_n^{-1}]^{1/2}
\\ [2ex]
\wt\cQ_n=\cQ_n-
 \cD_{n+2}\cU_n\cU_{n+2}^{-1}+\cD_n\cU_{n-2}\cU_n^{-1}
 \,.
\end{array}
\right.
\end{equation}
The matrices $\cG_n$ and $\cR_n$ represent in this case
 the ``potential differences"
 $\cG_n=\wt\cD_n-\cD_n$, $\cR_n=\wt\cQ_n-\cQ_n$
 which are expressed in terms of known quantities only.
Solutions of (\ref{tdnrn}) are found with the aid of
(\ref{Lpsin}), $\cA_n$ and $\cB_n$ given in (\ref{Anp})--(\ref{Bnp}):
\begin{equation}\label{wtpsin}
\wt\Psi_n=\cD_{n+2}^{1/2}(\cU_{n}\cU_{n+2}^{-1})^{1/2}
(\Psi_{n+2}-\cU_{n+2}\cU_n^{-1}\Psi_n) \,.
\end{equation}
Thus, we have solved the eigenvalue problem for a Hamiltonian
$\cH_1=\cH_0+\cV$ having
in the basis $|n,\alpha \rangle $
 an interaction in the form of the
block-tridiagonal matrix
\begin{equation}\label{interact}
\la k\beta |\cV |n\alpha\ra = G_{\beta ,\,\alpha ,n}\delta_{n-2,k}+
G_{\beta ,\,\alpha ,n+2}\delta_{n+2,k}+ R_{\beta ,\, \alpha,
n}\delta_{n,k}.
\end{equation}
 $G_{\beta ,\,\alpha ,n}$ and $R_{\beta ,\, \alpha, n}$ are
entries for $\cG_n$ and $\cR_n$ respectively, and  are found
from (\ref{dnfi})--(\ref{wtDp}).
 Hence, using (\ref{Dmatr})
 and the notation
 $H_{n}(\sqrt{2\lambda_1})=H_{n,1}$ and
 $H_{n}(\sqrt{2\lambda_2})=H_{n,2}$,
 we find 
$$
\cG_n=- \tfrac 14\sqrt{n(n-1)}\,\,\cI\,+
 \left(
 \begin{array}{cc}
 a_+ & a_- \\
 a_- & a_+
 \end{array}
 \right) \quad \mbox{ and}\quad 
\cR_n=
 \left(
 \begin{array}{cc}
 b_+ & b_- \\
 b_- & b_+
 \end{array}
 \right)
$$
with
\begin{equation}
 \begin{array}{l} 
a_{\pm}=\tfrac{\sqrt{n(n-1)}}{8}
 \left[
 \sqrt{\tfrac{H_{n+2,1}H_{n-2,1}}{(H_{n,1})^2}}\pm 
  \sqrt{\tfrac{H_{n+2,2}H_{n-2,2}}{(H_{n,2})^2}} \,\right]
 \\ [3ex]
 b_{\pm}=\tfrac{n(n-1)}{4}
 \left[
 \tfrac{H_{n-2,1}}{H_{n,1}}\pm 
 \tfrac{H_{n-2,2}}{H_{n,2}}
 \right] -
 \tfrac{(n+1)(n+2)}{4}
 \left[
 \tfrac{H_{n,1}}{H_{n+2,1}}\pm
 \tfrac{H_{n,2}}{H_{n+2,2}}
 \right].
\end{array}
\label{a12p}
\end{equation}
Note that the quantities $H_{n,1}$ and $H_{n,2}$ are real for $n$
even and purely imaginary for $n$ odd and therefore
$\cG_n$ and $\cR_n$ are real in either case.

 Let us find now the asymptotic behaviour of these quantities with
 respect to the discrete variable $n$.
 For this purpose we need an
 appropriate asymptotic expression for the Hermite polynomials, which can
be found with the help of the relationship between Hermite and Laguerre
polynomials, $H_{2n}(z)=(-1)^n2^{2n}n!L_n^{-1/2}(z^2)$, and
 using the
  series expansion for the Laguerre polynomials
 (see \cite{Bateman}, Eqs. 10.15(4), 10.15(5)) in which we keep only the
two first terms and consider $z^2<0$, $\mbox{Im} (z)>0$.
 We then have
\begin{equation}
 L_n^{-1/2}(z^2)=\frac{(2n-1)!!}{2^{n+1}n!}\,e^{z^2/2-2iz\sqrt{n}}
 \left[ 1-\tfrac{z^2}{16n}+O(n^{-3/2})\right]
\end{equation}
and
\begin{equation}\label{assH}
 H_{2n}(\sqrt{2\lambda })= \frac{(-2)^n \, (2n-1)!!}2
 \,e^{\lambda -2i\sqrt{2\lambda n}}
 \left[1-\tfrac{\lambda }{8n}+O(n^{-3/2})\right] .
\end{equation}
From (\ref{a12p}) we obtain
$a_+= n/4+ 1/8 -1/(32n) +O(1/n^2), 
b_+=1/2+O(1/n^2)$,
$a_-= O(1/n^2)$, and $b_-=O(1/n^2)$.
 The matrices $\cG_n$ and $\cR_n$ become diagonal
 up to terms of order $1/n^2$.
 Moreover, up to quadratic  terms in $1/n$,
 $a_+$ coincides with $d_{n+1}$  and $(q_n+b_+)$ with $q_{n+1}$.
This  means that for
 $n\to \infty$ the Darboux transformed
 potentials $\wt\cD_n$ and $\wt\cQ_n$ coincide with the initial ones,
up to the shifting $n\to n+1$.
 This is  the
discrete multichannel analogue of a property of
the usual Darboux  transformation saying that usually  the potential
difference vanishes asymptotically.

Now we will estimate the asymptotics of  solutions of the transformed
discrete eigenvalue problem.
To do it we will assemble  the column-vectors $\Psi_n^+(E_1)$ and
$\Psi_n^-(E_2)$ in the matrix
$\Xi_n$, which is a matrix
solution of the initial equation corresponding to the
diagonal matrix eigenvalue with entries $E_1$ and $E_2$. The matrix
$\wt\Xi_n:=(\wt\Psi_n^+(E_1),\wt\Psi_n^-(E_2))$ (the columns
$\wt\Psi_n^\pm(E_{1,2})$ are given by (\ref{wtpsin}) with the
replacement of $\Psi_n$ by $\Psi_n^\pm{(E_{1,2})}$ respectively)
is the
corresponding matrix solution to the transformed equation
\begin{equation}
\wt\Xi_n=\cP\Xi_n\,,\quad
\cP=\cD_{n+2}^{1/2}(\cU_{n+2}\cU_{n}^{-1})^{1/2}
(\cU_{n}\cU_{n+2}^{-1}\Xi_{n+2}\Xi_n^{-1}-\cI).
\end{equation}
Then, using the same asymptotics for the Hemite polynomials
(\ref{assH}), we obtain
\begin{equation}
\cP=
\left(
\begin{array}{cc}
p_1&q \\
q&p_2
\end{array}
\right),\ 
\begin{array}{l}
p_{1,2}=\sqrt{E_{1,2}}-\frac 12
(\sqrt{\lambda_1}+\sqrt{\lambda_2})+o(n^{-1/2})
\\[.5em]
q=\frac 12 (\sqrt{\lambda_2}-\sqrt{\lambda_1}\,)+o(n^{-1/2}).
\end{array}
\end{equation}
We see that for $n\to \infty$ the matrix $\cP$  does not depend on the
variable $n$, a result which agrees with the fact that the potential difference
vanishes asymptotically. This means, in particular, that if we choose  the
initial waves propagating only in one direction in both channels,
the same  asymptotic behavior  will be exhibited by the solutions of the transformed
equation. This property is precisely a two-channel analogue of the
transparency for the transformed potential we have obtained.

\sect{Conclusion}

In this paper we have introduced
a difference intertwining operator for discrete Schr\"odinger
equations with operator-valued coefficients
and then we studied some of its basic properties, such as
the factorization and the possibility to
establish a one-to-one correspondence between the spaces of
solutions of the initial and transformed equations.
This approach allowed us to get new exactly solvable eigenvalue
problems for block-tridiagonal matrices.
By working out the particular example of a two channel Schr\"odinger
equation, we have found a wide class of exactly solvable two-channel
potentials represented by block-tridiagonal matrices.

One of the possible applications of the method we have developed may be in
describing the scattering of composite particles, such as nucleons or
atoms, in the frame of the multichannel scattering theory \cite{Newton}.
This possibility is supported by the known applications of the
$J$-matrix method for describing the one-channel scattering \cite{J,VBA}.
Precisely in this line, the
next step of this work will be to find the transformation of the
scattering data for the spectral problem on a semi-axis.

\section*{Acknowledgments}

This work is supported by the Spanish MCYT
(BFM2002-03773), Junta de Castilla y Le\'on (VA085/02), and MECD
(BFS grant SAB2000-0240). We thank Professor M~L~Glasser for careful
reading of the manuscript and useful suggestions.

\end{document}